\documentclass[a4paper,twocolumn,11pt,noarxiv]{quantumarticle}
\pdfoutput=1
\usepackage[utf8]{inputenc}
\usepackage[english]{babel}
\usepackage[T1]{fontenc}
\usepackage{amsmath}
\usepackage{hyperref}

\usepackage[numbers]{natbib}

\usepackage{tikz}
\usepackage{lipsum}

\usepackage[utf8]{inputenc}
\usepackage[english]{babel}
\usepackage[T1]{fontenc}
\usepackage{amsmath}
\usepackage{hyperref}

\usepackage{braket}

\definecolor{cset-aps-blueberry}{RGB}{28,128,158}
\definecolor{cset-aps-blue}{RGB}{46,44,184}
\definecolor{cset-aps-turquoise}{RGB}{0,67,88}
\definecolor{cset-aps-limegreen}{RGB}{190,219,67}
\definecolor{cset-aps-green}{RGB}{31,138,112}
\definecolor{cset-aps-yellow}{RGB}{255,225,25}
\definecolor{cset-aps-orange}{RGB}{253,116,0}
\definecolor{cset-aps-red}{RGB}{219,0,43}

\definecolor{blau}{HTML}{1575B9}
\definecolor{hellblau}{HTML}{65B7EF}
\definecolor{rot}{HTML}{E31B0A}
\definecolor{hellrot}{HTML}{FC6761}
\definecolor{gruen}{HTML}{25A131}
\definecolor{lila}{HTML}{2E2CB8}
\definecolor{grau}{HTML}{A6A6A6}

\usepackage{hyperref}
\hypersetup{
    colorlinks=true,
    linkcolor={cset-aps-red},
    linkbordercolor={cset-aps-red},
    filecolor={cset-aps-orange},
    filebordercolor={cset-aps-orange},
    citecolor={cset-aps-blue},
    citebordercolor={cset-aps-blue},
    urlcolor={cset-aps-green},
    urlbordercolor={cset-aps-green},
    menucolor={cset-aps-limegreen},
    menubordercolor={cset-aps-limegreen},
    breaklinks=true,
    pdfborderstyle={/S/U/W 2},
    pdfpagemode=UseOutlines,
    pdfstartpage={1},
}

\usepackage{dsfont}

\newcommand{\tr}{\operatorname{tr}}

\newcommand{\ii}{\text{i}}
\newcommand{\dd}{\text{d}}

\newcommand{\id}{\hat{\mathds{1}}}

\newcommand{\affTUDaS}{
    \href{https://ror.org/05n911h24}{Technische Universit{\"a}t Darmstadt},
    Fachbereich Physik,
    Institut f{\"u}r Angewandte Physik,
    Schlo{\ss}gartenstr. 7,
    64289 Darmstadt,
    Germany
}

\begin{document}

\title{Entanglement and Its Verification: A Tutorial on Classical and Quantum Correlations}

\author{Enno Giese}
\affiliation{\affTUDaS}
\orcid{0000-0002-1126-6352}

\maketitle

\begin{abstract}
    Entanglement, a defining property of quantum mechanics in which two physical subsystems cannot be seen as independent entities, challenges our everyday experience and classical intuition.
    However, only such strong quantum correlations enable quantum technologies, including quantum computing or communication, while revealing the limits of our classical worldview by violating local realism.
    Given its importance in modern quantum science, we present this tutorial addressing the questions:
    What is entanglement, how does it differ from classical correlations, and how can it be experimentally verified?
    Using celebrated examples, such as Schrödinger's cat, we highlight the distinction between classical and quantum correlations and illustrate the definition of entangled and separable states.
    We review entanglement criteria by discussing Heisenberg-type uncertainty relations for continuous variables and the CHSH inequality for discrete systems.
    Focusing on concepts of quantum correlations and operational entanglement witnesses, we provide accessible tools and illustrative examples aimed at demystifying entanglement for a broad readership.
\end{abstract}

\section{Introduction}
\label{sec:Intro}

\begin{figure}
    \includegraphics[width=\linewidth]{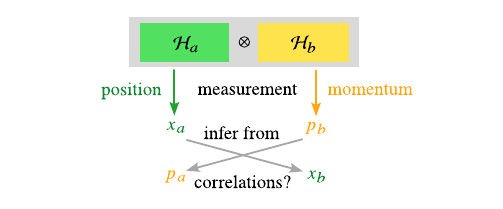}
    \caption{Visualization of the idea behind the EPR paradox~\cite{Einstein35}.
    Independent measurements of observables associated with subsystem $a$ or $b$, each described by elements of the associated Hilbert space $\mathcal{H}_j$ can always be performed with absolute certainty.
    The EPR state $\ket{\psi_\text{EPR}}= \int\! \dd p \, \psi(p) \ket{p}_a \otimes \ket{-p}_b$ is a valid element of the joint Hilbert space and displays strong anticorrelations of the momenta between both subsystems.
    This raises the conceptual question: given these perfect anticorrelated momenta and the ability to measure $x_a$ and $p_b$ with certainty, is it possible to infer with help of the correlation between $p_b$ and $p_a$ both $x_a$ and $p_a$ with certainty?
    }
    \label{fig:EPR-paradox}
\end{figure}

One of the key developments in quantum mechanics was the realization that physical observables are represented by self-adjoint (Hermitian) operators, whose eigenvalues correspond to the possible outcomes of measurements.
Physical systems, in turn, are described by elements of a Hilbert space called states.
The fact that operators may not commute has profound consequences for the precision of quantum measurements:
Two quantities described by noncommuting observables cannot be measured simultaneously with absolute certainty.
Indeed, a nonvanishing commutator between two observables implies uncertainty relations~\cite{Robertson29,Schroedinger30}, among them the most renowned Heisenberg uncertainty relation~\cite{Heisenberg27}
\begin{equation}
\label{eq:Heisenberg_uncertainty0}
    \Delta x \Delta p \geq \hbar /2 ,
\end{equation}
where the product of the standard deviations of position $\Delta x$ and momentum $\Delta p$ is bounded from below by half of Planck's constant $\hbar$.

In their discussion of the nature of a physical reality in quantum mechanics, A. Einstein, B. Podolsky, and N. Rosen (EPR) devised a gedankenexperiment~\cite{Einstein35} illustrated in Fig.~\ref{fig:EPR-paradox}:
A physical system may consist of two (potentially spatially separated) subsystems $a$ and $b$, each described by a Hilbert space $\mathcal{H}_\ell$.
These two subsystems can, for example, be two particles that might interact, but which are seen in classical physics as separate entities.
All states $\ket{\psi}$ describing the entire system must be elements of the joint Hilbert space, namely $   \ket{\psi} \in \mathcal{H}_a  \otimes \mathcal{H}_b$.
However, it is still possible to perform measurements on each subsystem independently (sometimes referred to as local measurements), for example by determining the position $x_a$ of subsystem $a$ and the momentum $p_b$ of subsystem $b$.
Since the corresponding operators act on different Hilbert spaces, they commute, and thus there is no uncertainty relation between these two properties of different subsystems.

EPR proposed the quantum state $\ket{\psi_\text{EPR}}= \int\! \dd p \, \psi(p) \ket{p}_a \otimes \ket{-p}_b$, which is an allowed element of the joint Hilbert space.
This state describes a system in which the momenta of both subsystems are perfectly anticorrelated, that is, $p_a=-p_b$, and $|\psi(p)|^2$ is the joint momentum probability density, depending only on single parameter $p$.
The associated distribution cannot be separated into independent probability distributions for subsystems $a$ and $b$.
Exactly this insight is central to the definition of entanglement.

Because the state $\ket{\psi_\text{EPR}}$ describes two subsystems whose momenta  $p_a$ and $p_b$ are strongly anticorrelated, one might be led to conclude that a measurement of $p_b$ allows one to infer the value $p_a$ without without direct measuring subsystem $a$.
Thus, a measurement of $x_a$ and independently of $p_b$ would seem to determine both $x_a$ and indirectly $p_a$ with certainty, apparently contradicting the Heisenberg uncertainty relation~\eqref{eq:Heisenberg_uncertainty0}.
However, this reasoning overlooks the fact that inferring the momentum of one subsystem by measuring the other is also associated with an uncertainty~\cite{Reid09}.
Although these quantum correlations may seem paradoxical at first glance~\cite{Popper34}, they reveal aspects of quantum physics that are entirely absent in classical theory.

Hence, two quantum subsystems may not constitute independent entities and therefore cannot be described separately, but only by states that are elements of the joint Hilbert space.
Erwin Schrödinger noted that such states are associated with a joint probability amplitude and coined the term entanglement~\cite{Schroedinger1935a,Schroedinger1935b} to described these intertwined, correlated subsystems.
This kind of inseparability does not occur in classical physics~\cite{Horodecki09}, where two subsystems always remain independent entities, even when strongly correlated.

While not all quantum states are entangled, entanglement is a uniquely quantum phenomenon, unlike statistical correlations of classical systems.
Most strikingly, entanglment can manifest as nonlocal correlations~\cite{Brunner14}, that is, even if the two subsystems are spatially separated, giving rise to what A. Einstein~\cite{Einstein2005} famously called \emph{spooky action at a distance}~\cite{Boughn22}.
In many-body systems, interaction-induced entanglement is both ubiquitous and is essential for describing collective quantum behavior, phase transitions, and correlated quantum systems.
Indeed, intrinsic entanglement naturally arises in condensed matter systems but cannot always be harnessed as a resource.
In contrast, the controlled generation of entanglement enables its use as a resource~\cite{Chitambar19} for quantum technologies:
Entangling gates between quantum bits are crucial to algorithms that outperform classical counterparts in quantum computing~\cite{Jozsa03}.
Since measurements by potential eavesdroppers induce decoherence and destroy entanglement, eavesdropping becomes detectable in quantum communication~\cite{Eckert91}, enabling secure quantum key distribution~\cite{Acin06}.
Entangled quantum states used in quantum sensors and metrological applications~\cite{Giovannetti11,Pezze18} give rise to sensitivities below shot noise, surpassing the standard quantum limit.

As such, entangled states exhibit properties that cannot be explained by classically correlated systems~\cite{Paneru20}.
Notably, Bell~\cite{Bell64} tests~\cite{Freedman72,Fry76,Aspect82a,Aspect82b,Weihs98} demonstrate that entangled quantum systems can exhibit correlations between spatially separated systems unexplainable by any local hidden variable theory~\cite{Bell66,Brunner14}.
These foundational insights led to questioning the notion of local realism.
Hence, the detection and characterization of entanglement remains an active field of research.

Because it is fundamental to many quantum technologies, certifying entanglement is of major relevance to applied quantum science.
A variety~\cite{Horodecki09,Yang25} of mathematical and operational entanglement criteria~\cite{Nielsen01,Duan00,Simon00} have been developed, ranging from the positive partial transpose criterion~\cite{Peres96,Horodecki96,Horodecki97} to different entanglement witnesses~\cite{Guehne09}.
For a pedagogical approach, we also adopt an operational perspective to address the following questions:
What is entanglement, how does it differ from classical correlations, and how can it be experimentally verified?

In this spirit, we provide the formal definitions of separable and entangled states in Sec.~\ref{sec:Entanglement}, where we use Schr\"odinger's cat as an illustrative example to clarify the difference between classical and quantum correlations~\cite{ Ranade13,Paneru20}.
We then introduce two operational tools for entanglement verification: uncertainty-based entanglement criteria (the EPR-Reid criterion~\cite{Reid89}) for continuous-variable systems in Sec.~\ref{sec.Heisenberg-type_uncertainty}, and Bell-type inequalities (the CHSH criterion~\cite{Clauser69}), including a motivation for the combination of measured correlations, for two-level systems in Sec.~\ref{sec:CHSH_inequality}.
For completeness, we include detailed derivations of the EPR-Reid and CHSH criteria in Appendices~\ref{app_EPR_Reid_derivation} and \ref{sec:CHSH_derivation}, respectively and derive the Tsireslon bound in Appendix~\ref{sec:Tsreslon_derivation}.

\section{Entanglement}
\label{sec:Entanglement}

The modern definition of entangled states is a negative one: entangled states are not separable in a mathematical sense, meaning they cannot be written as a product of states of the subsystems.
In fact, a pure state $\ket{\psi}\in  \mathcal{H}_a  \otimes \mathcal{H}_b$ is called separable if and only if
\begin{equation}
    \ket{\psi}= \ket{\psi_a}_a \otimes \ket{\psi_b}_b, 
\end{equation}
where $\ket{\psi_\ell}_\ell \in \mathcal{H}_\ell$.
Pure states that cannot be written in this form are called entangled and exhibit quantum correlations between the two subsystems.

However, correlations are not exclusively a quantum feature and are also known from classical physics.
For example, the colors of the pair of socks the reader might be wearing right now are probably correlated, even though neither the feet nor the socks are quantum objects or display quantum features~\cite{Bell81}.
To account for such classical correlations within quantum theory, one typically describes quantum states by density matrices $\hat{\rho}$.

Hence, we generalize the notion of separability to mixed states.
A density matrix $\hat{\rho}$ acting on $ \mathcal{H}_a  \otimes \mathcal{H}_b$ is called separable if and only if
\begin{equation}
\label{eq:separable_mixed}
    \hat{\rho}= \sum\limits_i w_i \hat{\rho}_a^{(i)} \otimes \hat{\rho}_b^{(i)},
\end{equation}
where the subensemble $i$ is described by $\hat{\rho}_\ell^{(i)}$, which acts on the Hilbert space $\mathcal{H}_\ell$ associated with subsystem $\ell$.
Here, $0 \leq w_i \leq 1$ is the classical probability for subensemble $i$, normalized such that $\sum_i w_i = 1$.
While such states may also display correlations, in each subensemble $i$ the density matrices factor, that is, $\hat{\rho}_a^{(i)} \otimes \hat{\rho}_b^{(i)}$, so that both subsystems remain independent entities, just weighted with a classical probability $w_i$.

\subsection{Is Schrödinger's cat entangled?}

To make the formal definition of separable, that is, not entangled states more accessible, we turn to the illustrative gedankenexperiment of a cat put forward by E. Schrödinger in the same article where he coined the term entanglement~\cite{Schroedinger1935a}.
We will also see that correlations are not necessarily an indication of entanglement.

\begin{figure}
    \includegraphics[width=\columnwidth]{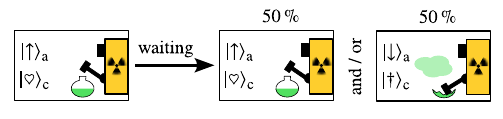}
    \caption{
    Schr\"odinger's cat~\cite{Schroedinger1935a} is locked alive ($\ket{\heartsuit}_\text{c}$) with an excited atom ($\ket{\uparrow}_\text{a}$) inside a sealed box, together with a Geiger counter and a vial of poison.
    After a period corresponding to the half-life of the atom, there is a chance of 50\,\% that the atom decays ($\ket{\downarrow}_\text{a}$).
    If a decay occurs, it is detected by the Geiger counter that destroys the vial, killing the 
    furball ($\ket{\dagger}_\text{c}$).
    }
    \label{fig:Schroedinger_cat}
\end{figure}

Let us assume that a cat, locked inside a box with an excited (radioactive) atom, can be described by two states:
dead ($\ket{\dagger}_{\text{c}}$) and alive ($\ket{\heartsuit}_{\text{c}}$).
The Hilbert space of the furball is then given by $\mathcal{H}_{\text{c}} = \text{span}\{\ket{\heartsuit}_{\text{c}}, \ket{\dagger}_{\text{c}}\}$.
While initially alive, the atom may decay via a quantum process.
This decay is detected by a Geiger counter, whose hammer, instead of striking the diaphragm of a loudspeaker, breaks a vial of poison, killing the cat.
After the half-life of the atom, the furry friend will be dead with a probability of 50\,\%, see Fig.~\ref{fig:Schroedinger_cat}.
Because this example was introduced to illustrate the inconceivability of quantum superpositions in our real-life experience, the discussion usually circles around the question of whether the cat is described by the state $(\ket{\heartsuit}_{\text{c}} + \ket{\dagger}_{\text{c}})/\sqrt{2}$, that is, simultaneously dead and/or alive after the half-life of the atom.

\begin{figure}
    \includegraphics[width=\columnwidth]{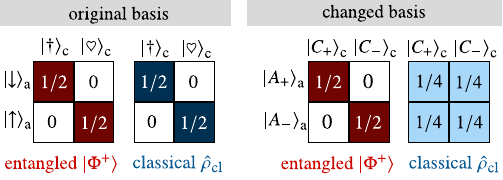}
    \caption{
    Correlation matrices between the states of the atom and the cat for both the entangled Bell state $\ket{\phi^+}$ and two classically correlated subsystems $\hat{\rho}_\text{cl}$.
    While in the original basis both states display strong correlations (left), they remain observable in a changed basis only for the entangled Bell state, while the classical state exhibits no correlations.
    The entries of each matrix represent the probability of detecting the corresponding joint outcome.
    }
    \label{fig:correlations-2levels}
\end{figure}

Here, we focus on a different aspect~\cite{Ranade13,Paneru20}:
In fact, we have two interacting subsystems, namely the cat and the atom described by the Hilbert space $\mathcal{H}_{\text{a}} = \text{span}\{\ket{\uparrow}_{\text{a}}, \ket{\downarrow}_{\text{a}}\}$, where $\ket{\uparrow}_{\text{a}}$ denotes the excited atom and $\ket{\downarrow}_{\text{a}}$ the decayed atom.
Because there is a causal connection between the atomic decay and the death of the cat, we always observe strong correlations between the furball being alive and the atom being excited, as well as the cat being dead and the atom having decayed.
In fact, the joint state
\begin{align}
\begin{split}
    \ket{\Phi^+} =  \frac{ \ket{\dagger}_{\text{c}} \otimes \ket{\downarrow}_{\text{a}} + \ket{\heartsuit}_{\text{c}} \otimes \ket{\uparrow}_{\text{a}}}{\sqrt{2}}
\end{split}
\end{align}
displays such correlations.
Indeed, $\ket{\dagger}_{\text{c}}$ is correlated with $\ket{\downarrow}_{\text{a}}$ and $\ket{\heartsuit}_{\text{c}}$ with $\ket{\uparrow}_{\text{a}}$, each with probability $1/2$.
No measurement will ever yield a result where $\ket{\dagger}_{\text{c}}$ is correlated with $\ket{\uparrow}_{\text{a}}$ or $\ket{\heartsuit}_{\text{c}}$ with $\ket{\downarrow}_{\text{a}}$.
The state $\ket{\Phi^+}$, often referred to as a Bell state, is entangled, since it cannot be written as a product of a state describing the cat and a state describing the atom, that is, $\left(\alpha_\text{c}  \ket{\dagger}_{\text{c}}  + \beta_\text{c} \ket{\heartsuit}_{\text{c}} \right) \otimes \left(\alpha_\text{a}  \ket{\downarrow}_{\text{a}}  + \beta_\text{a} \ket{\uparrow}_{\text{a}}  \right) \neq \ket{\Phi^+}$, with $|\alpha_\ell|^2 + |\beta_\ell|^2 = 1$.
The joint probability distribution shown in the correlation matrix in Fig.~\ref{fig:correlations-2levels} exhibits these strong correlations.
Therefore, we might be inclined to infer that Schrödinger's whiskered companion is entangled with the atom if we observe such correlations.

But, unfortunately, the observed correlations could also result from a classical process, similar to the example of the colors of socks given above.
Indeed, a classical mixture of a subensemle consisting of a dead cat and a decayed atom and a subensemble consisting of an alive cat and an excited atom, each with probability $1/2$, is encoded by the density matrix
\begin{equation}
    \hat{\rho}_\text{cl} = \frac{\ket{\dagger}_\text{c}\!\bra{\dagger}\otimes\ket{\downarrow}_\text{a}\!\bra{\downarrow}+\ket{\heartsuit}_\text{c}\!\bra{\heartsuit}\otimes\ket{\uparrow}_\text{a}\!\bra{\uparrow}}{2},
\end{equation}
which is by definition a separable state and not entangled.
As shown in Fig.~\ref{fig:correlations-2levels}, the correlation matrix of this mixed state is identical to the one observed for the Bell state.
Therefore, by simply checking whether the furball is dead or alive and whether the atom has decayed or is still excited, we cannot distinguish between the classical mixture and the entangled Bell state, and thus cannot determine whether Schrödinger's cat is entangled with the atom~\cite{Ranade13}.

As a side note, we observe that when no measurement is performed on the atom, the state of the cat is described by the partial trace over the atom $\tr_\text{a}\{ \ket{\Phi^+}\bra{\Phi^+} \}= (\ket{\dagger}_\text{c}\!\bra{\dagger}+\ket{\heartsuit}_\text{c}\!\bra{\heartsuit})/2=\tr_\text{a} \{\hat{\rho}_\text{cl}\}$ for both states.
Hence, the cat is always found in a classical mixture of being dead or alive, rather than a coherent superposition of being dead and alive~\cite{Rinner08}.

\subsection{Quantum vs. classical correlations}

However, if we perform a change of basis, we observe that these two joint probability distributions, that is, the correlation matrices, transform differently.
To illustrate this, we introduce a (local) change of basis for each subsystem, describing our system through the states
\begin{equation}
\label{eq:rotated_cat}
    \ket{C_\pm}_\text{c}= \frac{\ket{\heartsuit}_\text{c} \pm \ket{\dagger}_\text{c}}{\sqrt{2}}~ \text{ and }  ~  \ket{A_\pm}_\text{a}= \frac{\ket{\uparrow}_\text{a} \pm \ket{\downarrow}_\text{a}}{\sqrt{2}}.
\end{equation}
Such a change of the basis can be implemented through unitary (Hadamard) transformations, as formalized in Sec.~\ref{subsec:CHSH_motivation}.
Expressing both states in this new basis, we find
\begin{equation}
    \ket{\Phi^{+}}= \frac{ \ket{C_+}_{\text{c}} \otimes \ket{A_+}_{\text{a}} + \ket{C_-}_{\text{c}} \otimes \ket{A_-}_{\text{a}}}{\sqrt{2}},
\end{equation}
which remains entangled, that is, not separable and still strongly correlated.
The classically correlated density matrix
\begin{align}
\begin{split}
    \hat{\rho}_\text{cl} =& \frac{\ket{C_+}_\text{c}\!\bra{C_+}+\ket{C_-}_\text{c}\!\bra{C_-}}{2}\\& \quad \quad \quad \quad \otimes\frac{\ket{A_+}_\text{a}\!\bra{A_+}+\ket{A_-}_\text{a}\!\bra{A_-}}{2}    
\end{split}
\end{align}
is still separable.
However, in this basis we observe no correlations whatsoever for the classical state.
In particular, for the entangled state, $\ket{C_\pm}_\text{c}$ is always correlated with $\ket{A_\pm}_\text{a}$, and one never observes a combination $\ket{C_\pm}_\text{c}$ with $\ket{A_\mp}_\text{a}$.
For the classical state, no correlations persist and all combinations are possible, each with probability $1/4$.
These correlation matrices are shown on the right of Fig.~\ref{fig:correlations-2levels}.

Hence, we make the observation that while entanglement and classical correlations cannot be distinguished by measurements in a single basis, measurements in a changed basis reveal their difference and can be used to verify entanglement.
To distinguish quantum correlations, that is, entanglement, from classical ones, correlation measurements in at least two different bases are necessary.
Unluckily (or perhaps luckily) for the cat, up to now the measurement procedure for a living creature in the changed basis is unknown, in the sense that we cannot tell whether the furry friend is simultaneously dead and alive.
Therefore, we will never know if Schrödinger's whiskered companion is entangled with the atom.

\section{Heisenberg-type uncertainty relations}
\label{sec.Heisenberg-type_uncertainty}
Before we formalize how to distinguish between classical and quantum correlations in the context of two two-level systems such as Schrödinger's cat and an atom, we present in this section how to verify entanglement of continuous-variable systems.
As it turns out, there is a method that directly connects to the idea of the uncertainty relation within the EPR gedankenexperiment presented in Sec.~\ref{sec:Intro}, examining whether the observed correlations are in conflict with Heisenberg-type uncertainty relations.

\subsection{Entanglement criterion}

We therefore resort to position-momentum entanglement that was at the heart of the original idea of the EPR gedankenexperiment~\cite{Einstein35}.
We assume two subsystems $\ell = a, b$, each associated with elements of $\mathcal{H}_\ell = L^2$, the space of square integrable functions.
For each subsystem (or particle) $\ell$, the two conjugate continuous variables $\hat{x}_\ell$ and $\hat{p}_\ell$ acting on $\mathcal{H}_\ell$ describe the respective position and momentum.
In this case, we have the canonical commutation relation $[\hat{x}_\ell, \hat{p}_{\ell^\prime}] = \ii \hbar \delta_{\ell, \ell^\prime}$.
We consequently find the uncertainty relation~\cite{Heisenberg27}
\begin{equation}
    \Delta x_\ell \Delta p_\ell \geq | \langle [\hat{x}_\ell, \hat{p}_{\ell}] \rangle| /2 = \hbar /2
\end{equation}
for each subsystem.
However, these relations include no statement about correlations.
To quantify correlations, we introduce joint variables~\cite{Schroedinger1935b}, namely the relative position
\begin{subequations}
\label{eq:joint_variables}
    \begin{align}
        \hat{x}_- = &\frac{\hat{x}_a \otimes \id_b - \id_a \otimes \hat{x}_b}{\sqrt{2}}    
    \end{align}
and the mean momentum
    \begin{align}
       \hat{p}_+ = &\frac{\hat{p}_a \otimes \id_b + \id_a \otimes \hat{p}_b}{\sqrt{2}},
\end{align}
\end{subequations}
where the chosen normalization ensures that these observables correspond to axes rotated by $\pi/4$ in the joint position or momentum space, illustrated in Fig.~\ref{fig:correlations_continuous}.
Correlations in both position and momentum can be visualized by the width of the joint probability distribution along the antidiagonal and diagonal directions.
Thus, the variances $\Delta x_-^2$ of $\hat{x}_-$ and $\Delta p_+^2$ of $\hat{p}_+$ provide a quantitative measure of the strength of these correlations.

As a first point, we note that $[\hat{x}_-, \hat{p}_+] = 0$, so there is no fundamental uncertainty between these two joint variables, and the product $\Delta x_- \Delta p_+ \geq 0$ can take any value~\cite{Robertson29,Schroedinger30}.
In particular, observing $\Delta x_- \Delta p_+ < \hbar / 2$ is not in conflict with quantum mechanics, contrary to what one might naively expect from the discussion in Sec.~\ref{sec:Intro}.

\begin{figure}
    \includegraphics[width=\columnwidth]{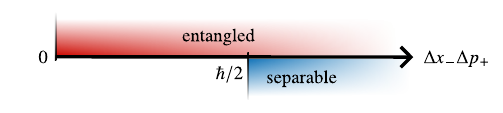}
    \caption{
    Visualization of the EPR-Reid criterion following Eq.~\eqref{eq:EPR-Reid}: For separable states, the product of uncertainties satisfies $\Delta x_- \Delta p_+ \geq \hbar /2$, a Heisenberg-type uncertainty relation between the position difference between two subsystems and the mean momentum. 
    Since $\hat{x}_-$ and $\hat{p}_+$ commute, entangled states are not constrained by this bound and can achieve $\Delta x_- \Delta p_+ < \hbar /2$.
    Hence, observing such a violation constitutes a direct verification of entanglement.
    }
    \label{fig:criterions_EPR_Reid}
\end{figure}

However, as derived in Appendix~\ref{app_EPR_Reid_derivation}, observing a value smaller than $\hbar/2$ implies that the two subsystems are necessarily entangled; thus, the observed correlations are genuinely quantum and cannot be explained by classical physics.
This result follows from the Cauchy-Schwarz inequality, the Heisenberg uncertainty relations for each subsystem, and the arithmetic mean-geometric mean inequality, all of which are rooted in the mathematical structure of quantum mechanics.
Summarizing these insights of Appendix~\ref{app_EPR_Reid_derivation}, we obtain the criterion~\cite{Reid89}
\begin{equation}
\label{eq:EPR-Reid}
    \Delta x_- \Delta p_+  \geq\begin{cases}
        0 & \text{for entangled states}\\
        \hbar /2 & \text{for separable states.}
    \end{cases}
\end{equation}
We find that a naive uncertainty relation between relative position and mean momentum holds for separable (classically correlated) systems, but can be violated by entangled states.
Thus, violating the naive Heisenberg-type uncertainty relation by a value smaller than $\hbar/2$ serves as a witness and proof of entanglement, known as the EPR-Reid criterion.
This criterion is illustrated in Fig.~\ref{fig:criterions_EPR_Reid}.

Quantum states can be prepared that exhibit strong position correlations, implying $\Delta x_-$ is small, while simultaneously displaying strong momentum anticorrelations, that is, $\Delta p_+$ is small, as illustrated in Fig.~\ref{fig:correlations_continuous}.
The essential features of this entanglement criterion align with Sec.~\ref{sec:Entanglement}:
it is necessary to measure correlations (implemented via joint variables along the diagonal and antidiagonal), and to perform measurements in different bases (such as position and momentum).
Moreover, this criterion shows that it is not required to probe the entire two-dimensional position and momentum spaces; it suffices to measure along specific directions, even if the direction of strongest correlation is initially unknown.
We note that related inequalities can serve as entanglement witnesses for continuous-variable systems, and the criterion presented here is just one illustrative example among many available in the literature.
Finally, this criterion can also be implemented to verify nonlocal entanglement: 
Since the narrow joint position distribution with strong correlations is releveant, but the absolute position in $(x_a, x_b)$ space does not matter, and the two subsystems can be measured far apart under conditions of space‑like separation.
In such a configuration, a violation of the inequality certifies entanglement between spatially separated subsystems, reproducing the essence of the EPR nonlocality argument.

\subsection{Example: Position-momentum entangled photons}

In the nonlinear process of spontaneous parametric down-conversion (SPDC), three different light fields are mixed within a medium exhibiting a second-order nonlinear susceptibility $\chi^{(2)}$, such as a nonlinear crystal.
Typically, a photon from a strong, classical laser, known as the pump field, is converted into two photons referred to as the signal and idler, for example distinguished by their polarization.
SPDC is a fundamental process in quantum optics and serves as the primary method for generating entangled photon pairs.
The key feature of SPDC is the nonlinear interaction of the three involved light fields, analogous to other mechanisms responsible for generating entanglement.

To study the correlations between the generated photons, one can place multi-pixel detectors in the two beams generated by a polarizing beam splitter, recording the transverse degrees of freedom of the signal and idler.
The correlation function is obtained by correlating, that is multiplying, the spatially resolved detection events from each camera, which can be positions so that the measurements are space-like separated to test for nonlocal correlations.
Alternatively, one may move slits in front of bucket detectors to record correlations pixel by pixel.
In our continuous-variable context, we assume that degenerate SPDC, i.\,e., signal and idler photons at the same wavelength.
To change the measurement basis, detectors can be placed in the far fields of the signal and idler (measuring transverse momenta) or the nonlinear medium can be imaged onto the detectors (measuring transverse positions in the near field).

The properties of the generated light are determined by the characteristics of the medium, encoded in the Fourier transform of the spatial profile of the nonlinearity along the longitudinal direction, denoted by the phase-matching function $\chi(p)$, and by the electric field $E(p)$ of the pump in the far field.
Although these quantities are generally tensors and vectors, their multiplication can be performed such that an effective scalar description is obtained.
In this scenario, the generated photon pair is described by the state
\begin{equation}
    \ket{\psi}\!\sim\!\! \int\!\!\dd p_a \dd p_b E(p_a+p_b) \chi\left(\!\frac{(p_a-p_b)^2}{2 \hbar k}\!\right) \ket{p_a}\otimes \ket{p_b},
\end{equation}
where $p_\ell$ denotes the transverse momentum of the signal ($a$) and idler ($b$) photons in a given direction, defined as the respective transverse component of their wave vector multiplied by $\hbar$.
Here, we have applied the paraxial approximation, neglected walk-off effects of the beams inside the medium, and used the modulus $k= 2\pi / \lambda$ of the pump's wave vector associated with its wavelength $\lambda$.
The modulus squared of the two-photon wave function, that is the integrand $E \chi$, can be interpreted as the joint probability distribution associated with the momenta of both photons.
This form closely resembles the state $\ket{\psi_\text{EPR}}$, which assumes perfect anticorrelation between the momenta of two subsystems and thus involves only one parameter.
However, the SPDC process generally allows for imperfect correlations, so whether correlations are observed depends on the specific shapes of the pump field and the nonlinearity.

\begin{figure}
    \includegraphics[width=\columnwidth]{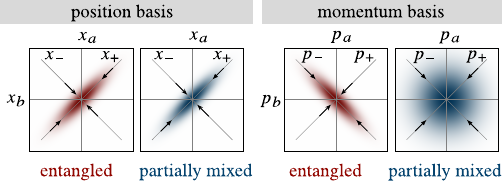}
    \caption{
    Joint probability distributions of continuous variables in the position basis (left) and momentum basis (right), where the darkness of the color represents the probability density in the joint space.
    The correlations in the entangled state’s position distribution can be inferred from the widths along the rotated coordinates  $ x_+ = (x_a + x_b)/\sqrt{2}$ and $ x_- = (x_a - x_b)/\sqrt{2}$. 
    Here, both the entangled and the partially mixed states exhibit the same position correlations, demonstrating that position correlations alone are insufficient to distinguish entanglement.
    In contrast, only the entangled state shows strong anticorrelations in the momentum basis, as seen along the corresponding rotated coordinates $p_+$ and $p_-$, whereas the partially mixed state exhibits no significant momentum correlations.
    }
    \label{fig:correlations_continuous}
\end{figure}

For a homogeneous crystal of length $L$, a coherent Gaussian pump beam with width $w$, and a phase-matching function approximated as Gaussian using the numerical factor $\alpha = 0.455$, the joint probability distribution for both momenta is given by~\cite{Law04,Walborn10}
\begin{align}
\begin{split}
    P(p_a,p_b)\sim &\,\exp\left[- \frac{2w ^2(p_a+p_b)^2}{\hbar^2}\right]\\
    &\quad \times  \exp\left[-\frac{\alpha L \lambda(p_a-p_b)^2}{4\pi\hbar^2 } \right].
\end{split}
\end{align}
The diagonal and antidiagonal directions in joint momentum space can be identified as $p_\pm = (p_a\pm p_b)/\sqrt{2}$.
Expressed in these variables, the probability distribution becomes
\begin{equation}
    P(p_a,p_b)\sim\exp\left[- \frac{p_+^2}{2 \Delta p_+^2}-  \frac{p_-^2}{2 \Delta p_-^2}\right]
\end{equation}
with the corresponding widths
\begin{equation}
    \Delta p_+^2 =  \hbar^2 /(8w^2) ~\text{ and } ~\Delta p_-^2= \hbar^2 \frac{ \pi  }{\alpha L \lambda}.
\end{equation}
An example of such a distribution is displayed in the left of Fig.~\ref{fig:correlations_continuous}.
For a broad Gaussian pump, the width $\Delta p_+$ is narrow, while for a short crystal, the width $\Delta p_-$ is broad, resulting in the detection of strongly anticorrelated momenta.
This behavior can be understood heuristically: if the pump beam approaches a plane wave, the generated photons propagate in opposite directions due to conservation of transverse momentum~\cite{Schneeloch16}.

By Fourier transforming the two-photon wave function from momentum to position representation and taking its modulus squared, we obtain the joint position distribution
\begin{equation}
    P(x_a,x_b)\sim\exp\left[- \frac{x_+^2}{2 \Delta x_+^2}-  \frac{x_-^2}{2 \Delta x_-^2}\right],
\end{equation}
where the widths along the diagonal and antidiagonal directions in joint position space are given by $\Delta x_\pm = \hbar / (2\Delta p_\pm)$.
See Fig.~\ref{fig:correlations_continuous} for an illustration.
As expected for a Fourier-limited state, the widths in position space are inversely proportional to the corresponding momentum widths.
Thus, the width of the position correlations is governed by the properties of the crystal and the pump field.
Strong correlations are observed in the same physical scenario where the momenta are anticorrelated.
This behavior can be interpreted heuristically as both photons being generated at the same transverse position~\cite{Schneeloch16}.

To verify entanglement, we consider the product
\begin{equation}
    \Delta x_-^2 \Delta p_+^2 =\frac{\hbar^2}{4} \frac{\Delta p_+^2}{\Delta p_-^2}= \frac{\hbar^2}{4} \frac{\alpha L \lambda}{8 \pi w^2},
\end{equation}
which provides bounds for the pump beam waist and the crystal length.
A short crystal and a large beam waist, corresponding to a pump approaching a plane wave, are favorable for entanglement verification.
In this regime, pronounced correlations and anticorrelations are observed in both position and momentum bases, respectively, enabling the observation~\cite{Ou92} of $\Delta x_- \Delta p_+ < \hbar /2$.
Because signal and idler photons have been spatially separated using a polarizing beam splitter, observing such a violation of the naive uncertainty relation certifies entanglement under nonlocal conditions can be sued to rule out local hidden variable theories.

Therefore, with the right choice of parameters, it is possible to generate and verify position-momentum entanglement.
However, this is not always guaranteed.
For example, a transversely incoherent pump field can be interpreted as a classical mixture of coherent pump beams, each characterized by a small effective beam waist determined by the transverse coherence length.
Such a mixed state, described by a density matrix, may exhibit correlations but not necessarily entanglement.
For pure states in momentum representation, the density matrix elements are given by $E(p_a+p_b)E^*(p_a'+p_b')$.
In the case of a mixed, partially coherent state, this product of electric fields must be replaced by the Fourier transform of the two-point correlation function
\begin{equation}
    \Gamma(x,x') \sim \exp\left[- \frac{x^2+x'^2}{4 w^2} -  \frac{(x-x')^2}{2 L_\text{c}^2 } \right],
\end{equation}
which describes the position correlations of an electric field of a Gaussian-Schell model beam of Gaussian width $w$ and transverse coherence length $L_\text{c}$.
Using such a beam does not affect the position correlations, since the intensity is evaluated at $\Gamma(x,x) \sim \exp\left[- x^2/(2 w^2)\right]$, corresponding to the modulus squared of the electric field, and we still observe $\Delta x_+ = \sqrt{2}w$.
However, the momentum distribution is affected, and the diagonal elements in the joint momentum space acquire a width that depends on the coherence length, so the state is no longer Fourier-limited.
Indeed, performing the explicit Fourier transform yields
\begin{equation}
    \Delta p_+^2= \frac{\hbar^2[1 + (2w/L_\text{c})^2]}{8 w^2} = \frac{1 + (2w/L_\text{c})^2}{4 \Delta x_+^2/ \hbar^2} 
\end{equation}
so that the joint momentum probability distribution broadens in this direction, as shown in Fig.~\ref{fig:correlations_continuous}.
The width in $p_+$ direction depends on the parameter $w/L_\text{c}$, and entanglement may not be verifiable~\cite{Giese18,Zhang19}.
To observe entanglement, it is ideal to have a coherence length much larger than the beam width $L_\text{c}\gg w$.
However, the inability to verify entanglement for such a pump does not necessarily imply that the photons are completely separable.

\section{CHSH inequality}
\label{sec:CHSH_inequality}

In the section above, we demonstrated that the detection of strong correlations between continuous variables in two conjugate bases can be used to construct an entanglement witness.
Such strong correlations in different bases can also be observed for two entangled two-level systems, such as Schrödinger's cat and the murdering atom introduced in Sec.~\ref{sec:Entanglement}.
In this section, we introduce a more measurement-based perspective, formalize the observations from Sec.~\ref{sec:Entanglement}, and present a criterion for entanglement of two two-level systems.

\subsection{Motivation}
\label{subsec:CHSH_motivation}

Instead of using the atom and cat example, where the measurement procedure for detecting a creature in a superposition of dead and alive is yet unknown, we introduce a more abstract and compact notation involving two two-dimensional Hilbert spaces $\mathcal{H}_\ell = \operatorname{span}\{ \ket{0}_\ell, \ket{1}_\ell \}$ for subsystems $\ell = a, b$.
We use the notation $\ket{i,j} = \ket{i}_a \otimes \ket{j}_b$.
These subsystems can be any physical two-level suystems, such as qubits, photon polarizations, atomic levels, spins, or motional states of trapped ions, among other implementations.
In this notation, the entangled Bell state and the classically correlated state are written as
\begin{equation}
\label{eq:Bell_class_states}
    \ket{\Phi^{+}}= \frac{\ket{11}+\ket{00}}{\sqrt{2}} ~
\text{and}~
    \hat{\rho}_\text{cl}= \frac{\ket{11}\!\bra{11}+\ket{00}\!\bra{00}}{2}.
\end{equation}

Next, we define a set of observables known as Pauli operators
\begin{subequations}
    \begin{align}
        \hat{\sigma}_{x,\ell}&= \ket{0}_\ell\!\bra{1}+\ket{1}_\ell\!\bra{0}\\
        \hat{\sigma}_{y,\ell}&=\ii \ket{0}_\ell\!\bra{1}-\ii\ket{1}_\ell\!\bra{0}\\
        \hat{\sigma}_{z,\ell}&= \ket{1}_\ell\!\bra{1}-\ket{0}_\ell\!\bra{0},
    \end{align}
\end{subequations}
which, together with the identity $\id_\ell$, form a basis for all linear operators acting on two-level systems.
These operators are Hermitian and unitary, and therefore involutory, meaning $\hat{\sigma}_{j,\ell}^2 = \ket{1}_\ell\!\bra{1}+\ket{0}_\ell\!\bra{0} = \id_\ell$.
This property also implies that they have two eigenvalues $\pm 1$, corresponding to the possible measurement outcomes.
Moreover, they satisfy the commutation relation $[\hat{\sigma}_{i,\ell},\hat{\sigma}_{j,\ell}] = 2 \ii \sum_k \epsilon_{ijk}\hat{\sigma}_{k,\ell}$, where $\epsilon_{ijk}$ is the antisymmetric Levi-Civita tensor.
Their product is given by $\hat{\sigma}_{i,\ell}\hat{\sigma}_{j,\ell}= \id_\ell \delta_{i,j} + \ii \sum_k \epsilon_{ijk} \hat{\sigma}_{k,\ell}$.
Most importantly, the Pauli operators are observables that can be implemented experimentally.
For example, $\hat{\sigma}_{z,\ell}$ can be measured by detecting the inversion of an atomic system.
In photonic platforms, these operators can be measured by analyzing photon polarization with different wave plate settings.
For spin systems, the Pauli operators correspond to projections of the spin along different directions, measured for example with the help of Stern-Gerlach setups.

To change the basis, we define the transformation $\hat{U}_\ell = (\hat{\sigma}_{x,\ell}+\hat{\sigma}_{z,\ell})/\sqrt{2}$, which is known in quantum information as the Hadamard gate.
Applying this transformation, we obtain the basis states
\begin{equation}
    \hat{U}_\ell \ket{1}_\ell = \frac{\ket{1}_\ell+\ket{0}_\ell}{\sqrt{2} }\text{ and } \hat{U}_\ell \ket{0}_\ell = \frac{\ket{1}_\ell-\ket{0}_\ell}{\sqrt{2}}
\end{equation}
in analogy to the new basis for the Schrödinger cat, see Eq.~\eqref{eq:rotated_cat}.
Hence, the states of interest in this basis are obtained from
\begin{subequations}
\begin{equation}
    \ket{\tilde{\Phi}^{+}} = \big(\hat{U}_a\otimes \hat{U}_b\big) \ket{\Phi^{+}}
\end{equation}
and
\begin{equation}
    \hat{\tilde{\rho}}_\text{cl}= \big(\hat{U}_a\otimes \hat{U}_b\big)^\dagger  \hat{\rho}_\text{cl} \big(\hat{U}_a\otimes \hat{U}_b\big).
\end{equation}    
\end{subequations}

To measure correlations we make a joint measurement of the inversion, implemented by the operator
\begin{equation}
    \hat{\sigma}_{z,a}\otimes \hat{\sigma}_{z,b} =\ket{11}\!\bra{11}+\ket{00}\!\bra{00} -\ket{01}\!\bra{01} - \ket{10}\!\bra{10}
\end{equation}
effectively summing the diagonal elements of the correlation matrix and subtracting the off-diagonal elements.
With the help of of Fig.~\ref{fig:correlations-2levels}, we find the measurement outcomes
\begin{subequations}
    \label{eq:sigmaz_sigmaz}
\begin{align}
    \bra{\Phi^{+}}\hat{\sigma}_{z,a}\otimes \hat{\sigma}_{z,b} \ket{\Phi^{+}} &=2 \times\frac{1}{2}-2 \times 0 \! =1\\
    \tr_{ab}\{ \hat{\rho}_\text{cl} \hat{\sigma}_{z,a}\otimes \hat{\sigma}_{z,b}\}&=2 \times\frac{1}{2}-2 \times 0\! =1
\end{align}
\end{subequations}
in the original basis, so that the result is unity for both the entangled and the classically correlated state, since their joint probability distributions are indistinguishable. 
In contrast, using the bottom panels of Fig.~\ref{fig:correlations-2levels}, we find in the changed basis
\begin{subequations}
\label{eq:sigmax_sigmax}
\begin{align}
    \bra{\tilde{\Phi}^{+}}\hat{\sigma}_{z,a}\otimes \hat{\sigma}_{z,b} \ket{\tilde{\Phi}^{+}} &= 2 \times \frac{1}{2}-2\times 0 \! =\!1\\
    \tr_{ab}\{ \hat{\tilde{\rho}}_\text{cl} \hat{\sigma}_{z,a}\otimes \hat{\sigma}_{z,b}\} &= 2\times \frac{1}{4}-2 \times \frac{1}{4} \! =\! 0
\end{align}
\end{subequations}
and observe a clear difference.

Because $\ket{\tilde\psi}_\ell = \hat{U}_\ell \ket{\psi}_\ell$,  the change of basis can be implemented by changing the observable.
Specifically, 
\begin{equation}
    _\ell\bra{\tilde\psi} \hat{\sigma}_{z,\ell}\ket{\tilde \psi}_\ell= {_\ell\!\bra{\psi}} \hat{U}_\ell^\dagger\hat{\sigma}_{z,\ell}\hat{U}_\ell\ket{ \psi}_\ell = {_\ell\!\bra{\psi}} \hat{\tilde{\sigma}}_{z,\ell}\ket{ \psi}_\ell ,
\end{equation}
where we define $ \hat{\tilde{\sigma}}_{z,\ell} =  \hat{U}_\ell^\dagger\hat{\sigma}_{z,\ell}\hat{U}_\ell$.
For the Hadamard transformation, we have $    \hat{\tilde{\sigma}}_{z,\ell}=  ( \hat{\sigma}_{x,\ell}+ \hat{\sigma}_{z,\ell}) \hat{\sigma}_{z,\ell}( \hat{\sigma}_{x,\ell}+ \hat{\sigma}_{z,\ell})/2$
so that we find
\begin{equation}
    \hat{\tilde{\sigma}}_{z,\ell}=   \hat{\sigma}_{x,\ell} + \frac{ \hat{\sigma}_{z,\ell}  +  \hat{\sigma}_{x,\ell} \hat{\sigma}_{z,\ell} \hat{\sigma}_{x,\ell}}{2} = \hat{\sigma}_{x,\ell},
\end{equation}
where we used $\hat{\sigma}_{j,\ell}^2  = \id_\ell$ in the first step, and identified $\hat{\sigma}_{z,\ell} \hat{\sigma}_{x,\ell} = \ii \hat{\sigma}_{y,\ell}$ as well as $\hat{\sigma}_{x,\ell} \hat{\sigma}_{y,\ell} = \ii \hat{\sigma}_{z,\ell}$ in the second step, due to the antisymmetric nature of the Levi-Civita symbol.
Hence, changing the basis corresponds experimentally to measuring $\hat{\sigma}_{x,\ell}$ instead of $\hat{\sigma}_{z,\ell}$.

To combine the measurements in both bases, we define the operator
\begin{equation}
    \hat{B}= \sqrt{2} (\hat{\sigma}_{z,a}\otimes \hat{\sigma}_{z,b}+\hat{\sigma}_{x,a}\otimes \hat{\sigma}_{x,b}),
\end{equation}
where the prefactor is a conventional choice that is motivated below. 
This operator sums correlation measurements performed in the original and in the changed basis.
To calculate the expectation value for the two states from Eq.~\eqref{eq:Bell_class_states}, we use the results obtained above in Eqs.~\eqref{eq:sigmaz_sigmaz} and \eqref{eq:sigmax_sigmax} to find
\begin{equation}
\label{eq:B_examples}
    |\langle \hat{B}\rangle| = 
\begin{cases}
 \sqrt{2} (1+1)= 2 \sqrt{2} &>2 ~ \text{ for } \ket{\Phi^{+}}\\
 \sqrt{2} (1+0)= \sqrt{2} &\leq 2 ~  \text{ for } \hat{\rho}_\text{cl}.
\end{cases}
\end{equation}
In the next section, we generalize this operator and connect it to an entanglement witness, where the bound of two is used to distinguish entangled from separable states.

\subsection{Entanglement criterion}

Because the optimal change of the basis, where classical correlations of an arbitrary mixed state are minimized, is unknown, we generalize the operator $\hat{B}$ by replacing $\hat{\sigma}_{z,a} \rightarrow \hat{a}_1$ and $\hat{\sigma}_{x,a}\rightarrow \hat{a}_2$, where $\hat{a}_j$ are two arbitrary (local) measurements on subsystem $a$ described by Hermitian operators.
Similarly, we generalize the measurements on subsystem $b$ by replacing $\hat{\sigma}_{z,b} \rightarrow (\hat{b}_1- \hat{b}_2)/\sqrt{2}$ and $\hat{\sigma}_{x,b} \rightarrow (\hat{b}_1+ \hat{b}_2)/\sqrt{2}$, introducing two arbitrary (local) measurements $\hat{b}_j$.
The observables can be represented in the Pauli-operator basis and remain involutory with $\hat{a}_j^2 = \id_a$ and $\hat{b}_j^2 = \id_b$, so the possible measurement outcomes are always $\pm 1$.
With these replacements, we generalize $\hat{B}$ to
\begin{subequations}
\begin{equation}
\label{eq:CHSH_operator_difference}
    \hat{\mathcal{B}}=  \hat{a}_1 \otimes (\hat{b}_1 - \hat{b}_2) + \hat{a}_2 \otimes( \hat{b}_1 +  \hat{b}_2),
\end{equation}
which can also be written as
\begin{equation}
\label{eq:CHSH_operator}
    \hat{\mathcal{B}}=  \hat{a}_1 \otimes \hat{b}_1 -\hat{a}_1 \otimes \hat{b}_2 + \hat{a}_2 \otimes \hat{b}_1 +\hat{a}_2 \otimes \hat{b}_2.
\end{equation}
\end{subequations}
In this form, it is apparent that the operator corresponds to four different correlation measurements, which are added and subtracted accordingly.
We refer to $\hat{\mathcal{B}}$ as the CHSH operator~\cite{Horodecki09}, after J. Clauser, M. Horne, A. Shimony, and R. Holt, who introduced such measurements in the context of Bell inequalities.
It is important to emphasize that the measurements can be performed locally, that is, independently on each subsystem, and that there are two measurement settings per subsystem.

\begin{figure}
    \includegraphics[width=\columnwidth]{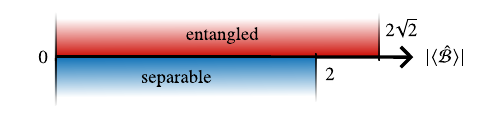}
    \caption{
    Visualization of the CHSH criterion following Eq.~\eqref{eq:CHSH_criterion}: For separable states, a Bell measurement will always lead to $|\langle\hat{\mathcal{B}}\rangle|\leq 2$.
    In contrast, entangled states can produce results $|\langle\hat{\mathcal{B}}\rangle|> 2$, serving as a whitness for entanglement.
    States for which the bound $|\langle\hat{\mathcal{B}}\rangle|= 2 \sqrt{2}$ is saturated are referred to as maximally entangled.
    }
    \label{fig:criterions_CHSH}
\end{figure}

Assuming separable states, we show in Appendix~\ref{sec:CHSH_derivation} using the triangle inequality, together with $\hat{a}^2_j = \id_a $ and $\hat{b}_j^2=\id_b$, that $|\langle\hat{\mathcal{B}}\rangle|$ is bounded by two.
In contrast, for arbitrary states we find in Appendix~\ref{sec:Tsreslon_derivation} using the operator norm and the Cauchy-Schwarz inequality that the bound is $2\sqrt{2}$.
Hence, we obtain the criterion
\begin{equation}
\label{eq:CHSH_criterion}
    0 \leq |\langle\hat{\mathcal{B}}\rangle| \leq \begin{cases}
         2 &\text{ for separable states}\\
         2 \sqrt{2} & \text{ for entangled states,}
    \end{cases}
\end{equation}
where the first line is known as the CHSH inequality, a specific manifestation of a Bell inequality, and the second line is called the Tsirelson bound~\cite{Cirelson80}.
Thus, measuring $|\langle\hat{\mathcal{B}}\rangle|> 2$ implies entanglement and a violation of the CHSH inequality, since the expectation values are bounded by two under the assumption of separable states.
If one even measures $|\langle\hat{\mathcal{B}}\rangle|= 2 \sqrt{2}$, saturating the Tsirelson bound, the state is called maximally entangled.
From Eq.~\eqref{eq:B_examples} we see that the Bell state $\ket{\Phi^+}$ is maximally entangled, while the classically correlated state $\hat{\rho}_\text{cl}$ does not violate the CHSH inequality.
We also show in Appendix~\ref{sec:Tsreslon_derivation} that in classical physics, where all observables commute, that is, $[\hat{a}_1,\hat{a}_2]=0=[\hat{b}_1,\hat{b}_2]$, the CHSH inequality always holds.
This insight tells us that noncommuting measurement settings are necessary to verifying entanglement.
The properties of the entanglement criterion~\eqref{eq:CHSH_criterion} are visualized in Fig.~\ref{fig:criterions_CHSH}.
Here we assume involuntary measurements that are implemented independently and locally for each subsystem, which can, in principle, be separated by a space-like distance.
Under these conditions, a violation of the CHSH inequality not only verifies entanglement but also rules out local hidden variable theories.

\subsection{Example: Polarization entanglement and Bell measurements}

Polarization-entangled photon pairs can also be generated by SPDC, where $\ket{1}$ corresponds, for example, to vertical polarization and $\ket{0}$ to horizontal polarization.
In such a setting, a Hadamard transformation corresponds to converting these states to diagonal and antidiagonal polarization, respectively.
In general, a change of polarization, and thus of the measurement basis, can be implemented using a half-wave plate set to a specific angle $\theta$ between the fast axis and a reference axis.
A wave plate thus realizes the angle-dependent observable
\begin{equation}
    \hat{\ell}_{\theta_\ell} = \hat{\sigma}_{z,\ell} \cos2\theta_\ell +\hat{\sigma}_{x,\ell} \sin2\theta_\ell 
\end{equation}
for each of the two photons, denoted by $\ell=a,b$.
As suggested by Eq.~\eqref{eq:CHSH_operator}, to verify entanglement with the CHSH inequality, it is necessary to measure four different correlations, implying two different wave plate settings $\theta_\ell$ per photon.
Typically, one chooses two different settings for subsystem $a$ and then scans $\theta_b$ to find the two settings for subsystem $b$ that maximize $ |\langle\hat{\mathcal{B}}\rangle| $.

\begin{figure}
    \includegraphics[width=\columnwidth]{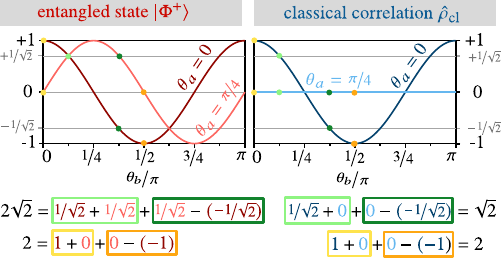}
    \caption{
    Observed correlations $\langle \hat{a}_{\theta_a}\otimes\hat{b}_{\theta_b}\rangle$ for a Bell test of polarization-entangled photons:
    Choosing the angle of the polarizer for subsystem $a$ to $\theta_a=0$ results identical behavior for both the entangled Bell state $\ket{\Phi^+}$ (left) and the classically correlated state $\hat{\rho}_\text{cl}$ (right) as a function of $\theta_b$, as shown by the dark traces.
    The lighter-colored traces correspond to the setting $\theta_a=0$, where the correlations differ between the two states.
    For a Bell test, four correlation measurements with appropriately chosen settings must be combined to evaluate the CHSH inequality.
    Using the set of green settings, maximal entanglement can be verified for $\ket{\Phi^+}$, while $\hat{\rho}_\text{cl}$ does not saturate the CHSH inequality (first line below the plot).
    In contrast, yellow settings allow $\hat{\rho}_\text{cl}$ to saturate the CHSH inequality, but fail to detect the entanglement in $\ket{\Phi^+}$ (second line below the plot).
    }
    \label{fig:Bell-measurement}
\end{figure}

Hence, we choose $\theta_a=0$, so that $\hat{a}_0= \hat{\sigma}_{z,a}$, and find the correlation
\begin{equation}
    \langle\hat{a}_0 \otimes \hat{b}_{\theta_b} \rangle = \cos2\theta_b   \langle\hat{\sigma}_{z,a} \otimes\hat{\sigma}_{z,b} \rangle+\sin 2\theta_b   \langle\hat{\sigma}_{z,a} \otimes\hat{\sigma}_{x,b} \rangle
\end{equation}
between signal and idler photons.
To evaluate the remaining expectation values, we consider as example the Bell state $\ket{\Phi^{+}}$ and the classically correlated state $\hat{\rho}_\text{cl}$ from Eq.~\eqref{eq:Bell_class_states}.
For these states, $\langle\hat{\sigma}_{z,a} \otimes\hat{\sigma}_{z,b} \rangle=1$, as can be directly seen from Fig.~\ref{fig:correlations-2levels} and the discussion in Sec.~\ref{subsec:CHSH_motivation}.
However, the expectation $\langle\hat{\sigma}_{z,a} \otimes\hat{\sigma}_{x,b} \rangle=0$ for both states so we find
\begin{equation}
    \langle\hat{a}_0 \otimes \hat{b}_{\theta_b} \rangle = \cos2\theta_b
\end{equation} 
for $\ket{\Phi^+}$ and $\hat{\rho}_\text{cl}$.
The correlations observed from scanning the angle $\theta_b$ for the setting $\theta_a=0$ are shown in Fig.~\ref{fig:Bell-measurement} by the dark traces.

In a second run of the experiment, we set the half-wave plate associated with Hilbert space $a$ to $\theta_a=\pi/4$, which implies $\hat{a}_{\pi/4}= \hat{\sigma}_{x,a}$, and hence we measure the correlations
\begin{align}
\begin{split}
    \langle\hat{a}_{\pi/4} \otimes \hat{b}_{\theta_b} \rangle = \cos&2\theta_b   \langle\hat{\sigma}_{x,a} \otimes\hat{\sigma}_{z,b} \rangle \\
    &+\sin 2\theta_b   \langle\hat{\sigma}_{x,a} \otimes\hat{\sigma}_{x,b} \rangle.
\end{split}
\end{align}
Similar to before, $\langle\hat{\sigma}_{x,a} \otimes\hat{\sigma}_{z,b} \rangle=0$ for both states considered, so the first summand vanishes.
However, $\langle\hat{\sigma}_{x,a} \otimes\hat{\sigma}_{x,b} \rangle=0$  for $\hat{\rho}_\text{cl}$ and $\langle\hat{\sigma}_{x,a} \otimes\hat{\sigma}_{x,b} \rangle=1$ for $\ket{\Phi^+}$, as already calculated in Sec.~\ref{subsec:CHSH_motivation}.
Therefore, we find  
\begin{equation}
    \langle\hat{a}_{\pi/4} \otimes \hat{b}_{\theta_b} \rangle = \begin{cases}
 \sin2\theta_b & ~ \text{ for } \ket{\Phi^{+}}\\
 0 & ~  \text{ for } \hat{\rho}_\text{cl}.
\end{cases}
\end{equation}
For the setting $\theta_a=\pi/4$ as well, $\theta_b$ is scanned, giving rise to the lighter traces in Fig.~\ref{fig:Bell-measurement}.

The predicted measurement results are depicted in Fig.~\ref{fig:Bell-measurement} for both the Bell state $\ket{\Phi^+}$ (left panel) and the classically correlated state $\hat{\rho}_\text{cl}$ (right panel).
For the entangled state, the combination  $(\theta_a,\theta_b) = \{(0,\pi/8),(0, 3\pi/8),(\pi/4,\pi/8),(\pi/4,3\pi/8)\}$ is optimal and leads to $|\langle \hat{B}\rangle|= 2 \sqrt{2}$, verifying that the Bell state is maximally entangled, saturating the Tsirelson bound.
In fact, these choices correspond to the change of basis through a Hadamard transformation discussed in Sec.~\ref{subsec:CHSH_motivation}, since $\hat{b}_{\pi/8}+\hat{b}_{3\pi/8}= \sqrt{2}\hat{\sigma}_{z,b}$ and $\hat{b}_{\pi/8}-\hat{b}_{3\pi/8}= \sqrt{2}\hat{\sigma}_{x,b}$.

In contrast, for this set of measurements, the classically correlated system yields $|\langle \hat{B}\rangle|=  \sqrt{2}\leq 2$, not even saturating the CHSH inequality.
However, we see in the figure that higher values of $ |\langle \hat{B}\rangle|$ can be achieved for $\hat{\rho}_\text{cl}$, specifically for $\theta_b = 0$ and $\theta_b= \pi/2$.
For this combination of settings, both the Bell state and the classically correlated state yield $ |\langle \hat{B}\rangle|= 2$, so that we do not verify entanglement for either.
While for the classically correlated state this choice is optimal and saturates the CHSH inequality, it is suboptimal for the entangled Bell state.
In general, measurement settings maximally violating the CHSH inequality depend on the entanglement of the state and can be found geometrically~\cite{Seiler21}. 
A key result is that the inability to verify entanglement in a given measurement does not imply that the state is not entangled.

\section{Conclusions}

Quantum entanglement is set apart from classical correlations by the persistence of strong correlations across different measurement bases, such that the two involved physical subsystems cannot be viewed as independent entities.
This feature enables operational verification of entanglement through measuring correlations with joint observables in varying bases.
We introduced two representative entanglement criteria: the uncertainty-relation-based EPR-Reid criterion for continuous variables~\cite{Reid89}, and the Bell-type CHSH inequality for discrete, two-level systems~\cite{Clauser69}.
As examples, we considered position-momentum entangled photon pairs that exhibit strong position correlations and momentum anticorrelations~\cite{Zhang19,Defienne19}, and we discussed the procedure for performing a Bell-type test~\cite{Freedman72,Fry76,Aspect82a,Aspect82b,Weihs98}.

While our focus was on operational detection of entanglement, a broader context includes foundational implications such as nonlocality and the exclusion of local realism and hidden variable theories by Bell tests.
The CHSH inequality constitutes just one approach among many~\cite{Popescu92,Brunner14} and ongoing research continues to refine loophole-free Bell tests~\cite{Shalm15,Giustina15,Hensen15} and develop alternative criteria in both theory and experiment.
To ensure locality, many of these tests~\cite{Aspect82b} use detectors that are separated by a space-like interval, so that the measurement processes are completed before any light-speed or classical signal could causally connect the two detectors. 

We illustrated two prominent operational criteria for bipartite entanglement, but many additional mathematical criteria~\cite{Nielsen01}, operational witnesses~\cite{Guehne09}, alternative uncertainty-relation-based inequalities~\cite{Teh14,Duan00,Simon00}, and other Bell-type tests exist~\cite{Brunner14}.
One can also extend uncertainty-based entanglement criteria to spatially distributed multi-particle systems, local measurements on separated atomic ensembles can reveal persistent entanglement, even when the subsystems are macroscopically distinct~\cite{Lange2018, Kunkel2018, Fadel2018}.
The examples presented here offer a conceptual, pedestrian introduction to entanglement, paving the way toward more advanced, system-specific, or technology-driven aspects or challenges posed by noise~\cite{Selim25}.
Viewed as a resource~\cite{Chitambar19}, quantifying entanglement using various measures~\cite{Adesso16} is especially important for multipartite~\cite{Jungnitsch11} and high-dimensional systems and in quantum technologies.
As efforts to generate entanglement advance across diverse platforms, including photons, ions, atoms, spin systems, macroscopic objects, and optomechanics, this tutorial serves as a starting point for exploring one of the most important and fascinating topics in quantum physics.

\begin{acknowledgements}
I thank J. Seiler, R. Fickler, A. Friedrich, as well as the Theoretical Quantum Optics group and students at TU Darmstadt for helpful comments and discussions.
\end{acknowledgements}

\onecolumn
\appendix

\section{EPR Reid entanglement criterion}
\label{app_EPR_Reid_derivation}

For the following derivation, our only assumption is that we are dealing with separable states $\hat{\rho}= \sum_i w_i \hat{\rho}_a^{(i)} \otimes\hat{\rho}_b^{(i)}$.
In each subensemble $i$, the density matrices associated with each subsystem factorize, that is, $\hat{\rho}_a^{(i)} \otimes \hat{\rho}_b^{(i)}$.
This allows us to take expectation values independently for each subsystem, as shown by
\begin{align}
    \tr_{ab}\big\lbrace \big(\hat{\rho}_a^{(i)}\otimes\hat{\rho}_b^{(i)} \big) \big(\hat{\mathcal{O}}_a \otimes \hat{\mathcal{O}}_b \big) \big\rbrace = \tr_a \big\lbrace \hat{\rho}_a^{(i)}\hat{\mathcal{O}}_a \big\rbrace \tr_b \big\lbrace \hat{\rho}_b^{(i)}\hat{\mathcal{O}}_b \big\rbrace ,
\end{align}
for any operators $\hat{\mathcal{O}}_\ell$ acting on $\mathcal{H}_\ell$.
We further define the expectation value in subensemble $i$ as $\langle \hat{\mathcal{O}}_\ell \rangle_i = \tr_\ell \lbrace \hat{\rho}_\ell^{(i)}\hat{\mathcal{O}}_\ell \rbrace$.

Since the entanglement criterion relies on the uncertainties of the joint operators from Eq.~\eqref{eq:joint_variables}, we begin by analyzing the variance of the relative position, defined as $\Delta x_-^2 = \langle \hat{x}_-^2 \rangle - \langle \hat{x}_- \rangle^2$.
To do this, we first need an expression for the second moment of $\hat{x}_-$, which requires evaluating the expectation value of $\hat{x}_-^2 = (\hat{x}_a^2 \otimes \id_b + \id_a \otimes \hat{x}_b^2 - 2 \hat{x}_a \otimes \hat{x}_b)/2$.
Using the definitions above, we find for a general separable state
\begin{equation}
    \langle\hat{x}_-^2\rangle = \frac{1}{2}\sum\limits_i w_i \left[\langle \hat{x}_a^2 \rangle_i + \langle \hat{x}_b^2 \rangle_i-2 \langle \hat{x}_a \rangle_i\langle \hat{x}_b \rangle_i\right].
\end{equation}
Introducing variances through $ \langle \hat{x}_\ell^2 \rangle_i = \Delta x_{\ell,i}^2 + \langle \hat{x}_\ell \rangle_i^2$ for each subensemble, and noting that $ \langle \hat{x}_a \rangle_i^2 +\langle \hat{x}_b \rangle_i^2 - 2 \langle \hat{x}_a \rangle_i\langle \hat{x}_b \rangle_i = 2\langle \hat{x}_- \rangle_i^2$, we obtain
\begin{equation}
    \langle\hat{x}_-^2\rangle = \frac{1}{2}\sum\limits_i w_i \left[\Delta x_{a,i}^2+\Delta x_{b,i}^2 \right] + \sum\limits_i w_i\langle \hat{x}_- \rangle_i^2
\end{equation}
for the second moment.
Therefore, with $\langle \hat{x}_- \rangle = \sum_i w_i \langle \hat{x}_- \rangle_i$, the variance can be written as
\begin{equation}
    \Delta x_-^2 = \sum\limits_i w_i \frac{\Delta x_{a,i}^2+\Delta x_{b,i}^2 }{2} +   \sum\limits_i w_i \langle \hat{x}_- \rangle_i^2 - \Big( \sum\limits_i w_i \langle \hat{x}_- \rangle_i \Big)^2 . 
\end{equation}

By artificially introducing $1 = \sum_j w_j$ before the second sum, and recalling the Cauchy-Schwarz inequality $\sum_j \alpha_j^2 \sum_i \beta_i^2 \geq (\sum_i \alpha_i \beta_i)^2$, we can identify the real parameters $\alpha_j = \sqrt{w_j}$ and $\beta_j = \sqrt{w_i} \langle \hat{x}_- \rangle_j$.
With this identification, it is apparent that the last two terms in the expression for the variance are always greater than or equal to zero.
They can therefore be omitted to obtain the lower bound
\begin{equation}
    \Delta x_-^2 \geq  \sum\limits_i w_i \frac{\Delta x_{a,i}^2+\Delta x_{b,i}^2 }{2} \geq \sum\limits_i w_i \Delta x_{a,i}\Delta x_{b,i},
\end{equation}
where, for the last step, we have applied the arithmetic mean-geometric mean inequality $\alpha^2 + \beta^2 \geq 2 \alpha \beta$, which holds for any real parameters $\alpha$ and $\beta$ and follows directly from $(\alpha-\beta)^2 = \alpha^2+\beta^2-2\alpha \beta \geq 0$.

By following exactly the same steps, we derive for the variance of the mean momenta the expression
\begin{equation}
    \Delta p_+^2 \geq \sum\limits_j w_j \Delta p_{a,j}\Delta p_{b,j},
\end{equation}
which characterizes the anticorrelations in the (conjugate) measurement basis.
This leads us to a first lower bound for the uncertainty product
\begin{equation}
    \Delta x_-^2\Delta p_+^2 \geq \Big(  \sum\limits_i w_i \Delta x_{a,i}\Delta x_{b,i}\Big) \Big( \sum\limits_j w_j \Delta p_{a,j}\Delta p_{b,j}\Big).
\end{equation}
Applying the Cauchy-Schwarz inequality once more, this product can be further bounded as
\begin{equation}
    \Delta x_-^2\Delta p_+^2 \geq \Big(  \sum\limits_i \sqrt{ w_i w_i\Delta x_{a,i} \Delta p_{a,i}\Delta x_{b,i}\Delta p_{b,i}}\Big)^2 .
\end{equation}
If we use the Heisenberg uncertainty relation for each subensemble, namely $\Delta x_{\ell,i} \Delta p_{\ell,i}\geq \hbar /2$, we find
\begin{equation}
\label{eq:EPR_Reid_app}
    \Delta x_- \Delta p_+ \geq  \sum\limits_i  w_i \hbar /2 = \hbar /2 
\end{equation}
with the help of $\sum_i w_i =1$.

Even though $\hat{x}_-$ and $\hat{p}_+$ commute and therefore $\Delta x_- \Delta p_+ \geq 0$ so that their uncertainties can, in principle, be made arbitrarily small, we obtain a Heisenberg-type uncertainty relation when we restrict ourselves to separable states.
In other words, if this Heisenberg-type uncertainty relation is violated in an experiment, that is, $\Delta x_- \Delta p_+ < \hbar /2$, the two systems must have been entangled, since separability was the only assumption used in the derivation.
In contrast, classically correlated (separable) systems will always satisfy the Heisenberg-type uncertainty relation given by Eq.~\eqref{eq:EPR_Reid_app}.

\section{CHSH inequality}
\label{sec:CHSH_derivation}
For any expectation value of the CHSH operator $\hat{\mathcal{B}}$ in the form of Eq.~\eqref{eq:CHSH_operator_difference}, which consists of a linear combination of four correlation measurements defined by operators $\hat{\ell}_j^2= \id_\ell$ acting on subsystems $\ell=a,b$ with two measurement settings $j=1,2$, the triangle inequality yields the bound
\begin{equation}
    |\langle\hat{\mathcal{B}}\rangle| \leq |\langle \hat{a}_1 \otimes (\hat{b}_1-\hat{b}_2)\rangle|+|\langle \hat{a}_2 \otimes (\hat{b}_1+\hat{b}_2)\rangle|.
\end{equation}
Assuming separable states $\hat{\rho}= \sum_i w_i \hat{\rho}_a^{(i)} \otimes\hat{\rho}_b^{(i)}$, and using the notation $\langle \hat{O}_\ell\rangle_i = \tr_\ell \lbrace \hat{\rho}_\ell^{(i)}\hat{\mathcal{O}}_\ell \rbrace$, we can take the expectation values of both subsystems independently in each subensemble.
This leads to the bound
\begin{equation}
    |\langle\hat{\mathcal{B}}\rangle| \leq \sum\limits_i w_i \left( |\langle \hat{a}_1\rangle_i| \, |\langle\hat{b}_1-\hat{b}_2\rangle_i|+|\langle \hat{a}_2\rangle_i| \,|\langle\hat{b}_1+\hat{b}_2\rangle_i|\right).
\end{equation}
Since all operators have eigenvalues $\pm 1$ that correspond to the two possible measurement outcomes, we observe $|\langle \hat{a}_j\rangle_i|\leq 1$ and therefore
\begin{equation}
    |\langle\hat{\mathcal{B}}\rangle| \leq \sum\limits_i w_i \left(|\langle\hat{b}_1-\hat{b}_2\rangle_i|+|\langle\hat{b}_1+\hat{b}_2\rangle_i|\right).
\end{equation}

Because the modulus appears in the equation above, we must consider two cases to bound the term in parentheses:
(i) If $|\langle \hat{b}_1\rangle_i|>|\langle \hat{b}_2\rangle_i|  $, we can apply the triangle inequality 
\begin{subequations}
\begin{equation}
    |\langle\hat{b}_1-\hat{b}_2\rangle_i|+|\langle\hat{b}_1+\hat{b}_2\rangle_i| \leq |\langle\hat{b}_1\rangle_i|-|\langle\hat{b}_2\rangle_i|+|\langle\hat{b}_1\rangle_i|+|\langle\hat{b}_2\rangle_i|
\end{equation}
so that the term is $\leq 2|\langle \hat{b}_1\rangle_i| $.
Since $|\langle \hat{b}_j\rangle_i|\leq 1 $, the whole term in parentheses is therefore  $\leq 2$.
(ii) Similarly, if $|\langle \hat{b}_1\rangle_i|<|\langle \hat{b}_2\rangle_i|  $, we use the bound
\begin{equation}
    |\langle\hat{b}_1-\hat{b}_2\rangle_i|+|\langle\hat{b}_1+\hat{b}_2\rangle_i| \leq -|\langle\hat{b}_1\rangle_i|+|\langle\hat{b}_2\rangle_i|+|\langle\hat{b}_1\rangle_i|+|\langle\hat{b}_2\rangle_i|
\end{equation}
\end{subequations}
so that the term is $\leq 2|\langle \hat{b}_2\rangle_i| $, and, using $|\langle \hat{b}_j\rangle_i|\leq 1 $, the whole term in parenthesis is again bounded by two.

As a consequence, together with $\sum_i w_i=1$, we find in both cases for separable states, which was the only assumption, the relation
\begin{equation}
      |\langle\hat{\mathcal{B}}\rangle| \leq 2
\end{equation}
referred to as CHSH inequality.
If this inequality is violated, our assumption of separable states must have been incorrect, so that the inequality may serve as a witness for entanglement.

entanglement

\section{Tsirelson bound}
\label{sec:Tsreslon_derivation}

Taking the square of the CHSH operator in form of Eq.~\eqref{eq:CHSH_operator_difference},  and using $(\hat{b}_1\pm\hat{b}_2)^2 = 2 \id_b \pm (\hat{b}_1\hat{b}_2+\hat{b}_2\hat{b}_1)$ as well as $(\hat{b}_1\mp\hat{b}_2)(\hat{b}_1\pm\hat{b}_2)=\pm[\hat{b}_1,\hat{b}_2]$, we find the identity
\begin{equation}
    \hat{\mathcal{B}}^2 =  4 \id_a\otimes\id_b + [\hat{a}_1,\hat{a}_2] \otimes [\hat{b}_1,\hat{b}_2].
\end{equation}
In classical physics, all observables always commute, so that we have $\hat{\mathcal{B}}^2 =  4 \id_a\otimes\id_b$ and, using the Cauchy-Schwarz inequality,
\begin{equation}
    |\langle\hat{\mathcal{B}}\rangle| \leq \sqrt{\rule{0pt}{1.75ex}\langle\smash{\hat{\mathcal{B}}}^2\rangle}= \sqrt{4}=2
\end{equation}
is the upper bound in classical physics, coinciding with the CHSH inequality.
This equation also demonstrates that it is necessary to use noncommuting observables for each system to verify entanglement.

If we rely on the operator norm $\Vert\hat{\mathcal{O}}\Vert = \sup_{\ket{\psi}} \bra{\psi} \hat{\mathcal{O}}\ket{\psi}$ induced by the inner product, where the supremum is taken over all $\ket{\psi}\neq 0$ normalized to unity, we can bound the norm of the squared CHSH operator by
\begin{equation}
    \Vert \hat{\mathcal{B}}^2 \Vert \leq 4 \Vert\id_a\otimes\id_b\Vert + \Vert[\hat{a}_1,\hat{a}_2]\Vert \, \Vert[\hat{b}_1,\hat{b}_2]\Vert \leq 4+4 =8,
\end{equation}
because $\Vert[\hat{a}_1,\hat{a}_2]\Vert \leq 2 \Vert\hat{a}_1\Vert \, \Vert\hat{a}_2\Vert \leq 2$ and similarly for the operators $\hat{b}_j$ acting on the Hilbert space associated with subsystem $b$.
Using the Cauchy-Schwarz inequality, we therefore find
\begin{equation}
    |\langle\hat{\mathcal{B}}\rangle| \leq \sqrt{\langle\rule{0pt}{1.75ex}\smash{\hat{\mathcal{B}}}^2\rangle} \leq \sqrt{\Vert \rule{0pt}{1.75ex}\smash{ \hat{\mathcal{B}}^2}\Vert} \leq \sqrt{8}= 2 \sqrt{2},
\end{equation}
where no assumptions about the separability of the states were made, so this bound is valid for both separable and entangled states.
This so-called upper Tsirelson bound~\cite{Cirelson80} is tight and holds for any measurement of $\hat{\mathcal{B}}$ in quantum mechanics.

\bibliographystyle{quantum}
\bibliography{literature}


\end{document}